\documentclass[proof]{WileyASNA-v1}

\articletype{Article Type}%

\received{19 July 2022}
\revised{* * 2022}
\accepted{* * 2022}

\begin{document}

\title{Multiwavelength astrophysics of the blazar OJ 287 and the project MOMO}

\author[1]{S. Komossa*}
\author[1]{A. Kraus}
\author[2]{D. Grupe}
\author[3]{M.L. Parker}
\author[4]{A. Gonzalez}
\author[4]{L.C. Gallo}
\author[5]{M.A. Gurwell}
\author[6]{S. Laine}
\author[1]{S. Yao}
\author[7]{S. Chandra}
\author[8]{L. Dey}
\author[9]{J.L. Gomez}
\author[10]{A. Gopakumar}
\author[11]{K. Hada}
\author[12]{D. Haggard}
\author[4]{A.R. Hollett}
\author[13]{H. Jermak}
\author[14,15]{S. Jorstad}
\author[1]{T.P. Krichbaum}
\author[16]{S. Markoff}
\author[13]{C. McCall}
\author[17]{J. Neilsen}
\author[18]{M. Nowak}

\authormark{S. KOMOSSA \textsc{et al}}

\address[1]{\orgdiv{MPIfR}, 
\orgaddress{\country{Bonn, Germany}}}
\address[2]{\orgdiv{Dept. of Physics, Geology, and Engineering Technology, NKU, Kentucky, USA}} 
\address[3]{\orgdiv{IoA, University of Cambridge, UK}}
\address[4]{\orgdiv{Department of Astronomy and Physics, Saint Mary's University, Halifax, Canada}}
\address[5]{\orgdiv{Center for Astrophysics | Harvard \&
Smithsonian, Cambridge, MA, USA}}  
\address[6]{\orgdiv{IPAC, Caltech, Pasadena, CA, USA}}
\address[7]{\orgdiv{Centre for Space Research,
North-West University, Potchefstroom, South Africa }}
\address[8]{\orgdiv{National Centre for Radio Astrophysics, TIFR, Pune, India}}
\address[9]{\orgdiv{Instituto de Astrofisica de Andalucia, Granada, Spain}}
\address[10]{\orgdiv{Department of Astronomy and Astrophysics, TIFR, Mumbai, India}}
\address[11]{\orgdiv{$^{11}$Department of Astronomical Science, The Graduate University for Advanced Studies (SOKENDAI), Tokyo, Japan}}
\address[12]{\orgdiv{Department of Physics, and McGill Space Institute, McGill Univ., Montréal, Canada}}
\address[13]{\orgdiv{Astrophysics Research Institute, Liverpool John Moores University, Liverpool, UK}}
\address[14]{\orgdiv{Institute for Astrophysical Research, Boston University, Boston, MA, USA}}
\address[15]{\orgdiv{Sobolev Astronomical Institute, St. Petersburg State University, Russia}}
\address[16]{\orgdiv{Anton Pannekoek Institute for Astronomy, University of Amsterdam, The Netherlands}}
\address[17]{\orgdiv{Villanova University, PA, USA}}
\address[18]{\orgdiv{Department of Physics, Washington University in St. Louis, MO, USA}}

\corres{*S. Komossa. \email{astrokomossa@gmx.de}}

\abstract{We are carrying out the densest and longest multiyear, multiwavelength monitoring project of OJ 287 ever done. The project MOMO (Multiwavelength Observations and Modelling of OJ 287) covers wavelengths from the radio to the high-energy regime. A few selected observations are simultaneous with those of the Event Horizon Telescope (EHT). MOMO aims at understanding disk-jet physics and at testing predictions of the binary black hole scenario of OJ 287. Here, we present a discussion of extreme outburst and minima states in context, and then focus on the recent flux and spectral evolution between 2021 and May 2022, including an ongoing bright radio flare.  Further, we show that there is no evidence for precursor flare activity in our optical--UV--X-ray light curves 
that would be associated with any secondary supermassive black hole (SMBH) disk impact and that was predicted to start as thermal flare on 2021 December 23.}


\jnlcitation{\cname{%
\author{Komossa S.}, 
\author{A. Kraus}, 
\author{D. Grupe}, 
\author{M.L. Parker},
\author{A. Gonzalez},
\author{L.C. Gallo},
\author{M.A. Gurwell},
\author{S. Laine},
\author{S. Yao},
\author{S. Chandra},
\author{L. Dey},
\author{J.L. Gomez},
\author{A. Gopakumar},
\author{K. Hada},
\author{D. Haggard},
\author{A.R. Hollett},
\author{H. Jermak},
\author{S. Jorstad},
\author{T.P. Krichbaum},
\author{S. Markoff},
\author{C. McCall},
\author{J. Neilsen}, and 
\author{M. Nowak}} (\cyear{2022}), 
\ctitle{Multiwavelength astrophysics of the blazar OJ 287 and the project MOMO}, \cjournal{Astronomische Nachrichten (XMM-Newton workshop 2022 proceedings)}, \cvol{2022;00:0--0}.}


\maketitle

\section{Introduction}\label{sec1}



Blazars are characterized by their powerful, collimated, long-lived jets of relativistic particles that often extend beyond the host galaxies. 
The jets are launched in the immediate environment of the supermassive black holes (SMBHs) at the blazars' centers \citep{Blandford2019}. The accretion disk -- jet interface represents one of the most extreme astrophysical environments where special and general relativistic effects play a major role in shaping the multiwavelength (MWL) radiation of these systems. 
Hence, blazars are excellent
laboratories for understanding matter under strong gravity and the disk-jet coupling.

\begin{figure*}
\centering
\includegraphics[clip, trim=1.0cm 5.3cm 1.5cm 9.0cm, width=12.5cm]{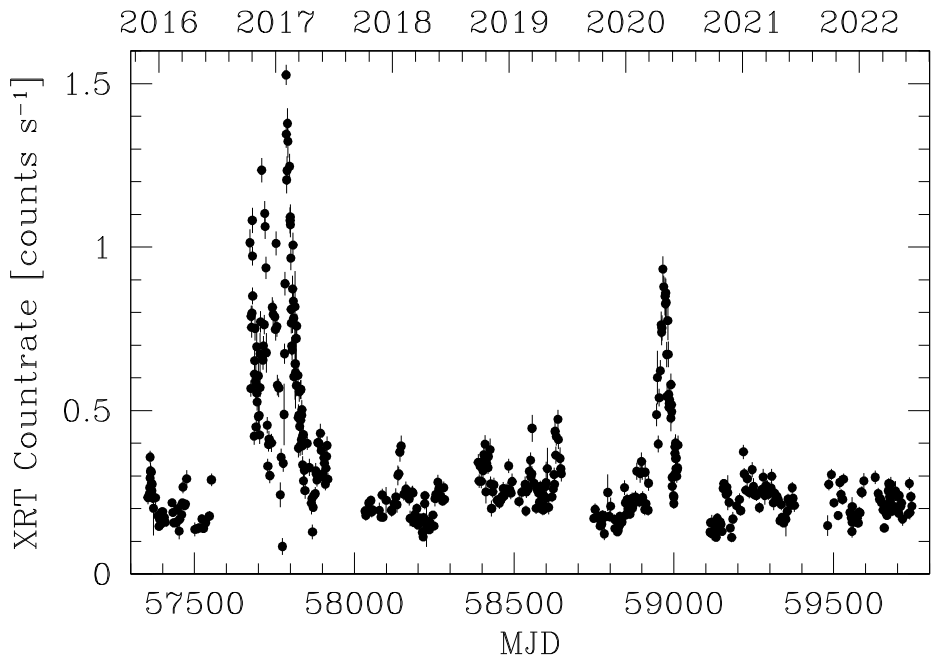}
    \caption{Swift X-ray telescope (XRT) light curve of OJ 287 between 2015 December and 2022 May. The large majority of the data were taken in the course of the MOMO program. Bright outbursts were detected in 2016--2017 and 2020 that came with a strong softening of the X-ray spectra.
    }
    \label{fig:light-CR-Swift}
\end{figure*}

OJ 287 is a nearby bright blazar of BL Lac type at redshift $z$=0.306,
and highly variable across the electromagnetic spectrum \citep[e.g.][]{Abdollahi2022, Komossa2022}. During epochs of outbursts 
(Fig. \ref{fig:light-CR-Swift})
it reveals a bright and supersoft X-ray emission component, making it one of the few blazars of its type that has contributions from both the synchrotron and inverse-Compton (IC) emission component present in its X-ray spectrum below 10 keV \citep{Komossa2020}. OJ 287 therefore offers the rare chance of studying the relation of both spectral components simultaneously in single-band observations.  

Among the main EHT targets, OJ 287 is the only source observed with XMM-Newton and EHT quasi-simultaneously in 2018 (M87 and SgrA* were unobservable with XMM-Newton at the epochs of EHT observations). 

Based on repeated bright optical outbursts in the 1970s to 1990s, it was suggested that OJ 287 hosts a binary SMBH \citep{Sillanpaa1988}. The amplitude of the brightest optical outbursts of OJ 287 decreased in recent years \citep[e.g.,][]{Dey2018, Laine2020, Komossa2021c}, making it more difficult to distinguish between blazar-driven activity and hypothetical binary-driven activity. 
A broad wavelength coverage and high-cadence monitoring are therefore important when testing model predictions. 

The project MOMO (Multiwavelength Observations and Modelling of OJ 287) was initiated in late 2015 \citep[][and references therein]{Komossa2021c}. It covers OJ 287 densely from the radio to the X-ray regime in dedicated observations with the Effelsberg telescope (between 1 and 40 GHz) and with the Neil Gehrels Swift observatory (Swift hereafter; in three optical bands, three UV bands, and in the 0.3--10 keV X-ray regime). Public data from the Submillimeter Array SMA \citep[$\sim$230 GHz;][]{Gurwell2007} and the Fermi satellite \citep[0.1--100 GeV;][]{Kocevski2021} are added. At selected epochs deep follow-up observations with XMM-Newton and NuSTAR and at other wavebands were taken, including during a few epochs of EHT observations of OJ 287. MOMO provides MWL timing information, spectra, and broad-band spectral energy distributions (SEDs) at all activity states of OJ 287. When new remarkable flux or spectral states of OJ 287 are discovered, the community is alerted rapidly in form of Astronomer's Telegrams \citep[e.g.][]{Komossa2021d}. 

Previous publications of results from the MOMO project focused on 1) our discovery with Swift of a bright X-ray--UV--optical outburst that started in 2016 \citep{Komossa2017},\citep{Komossa2020}, 2) our discovery with Swift of a second bright X-ray--UV--optical outburst in 2020 that we followed up with XMM-Newton and NuSTAR \citep{Komossa2020}, 3) a detailed analysis of all XMM-Newton X-ray spectra of OJ 287 including one quasi-simultaneous with EHT \citep{Komossa2021a}, 4) a study of the complete Swift optical--UV--X-ray light curve of OJ 287 covering all activity states of this blazar \citep{Komossa2021c}, and 5) selected multi-frequency Effelsberg radio observations of OJ 287 since 2015 \citep{Komossa2015, Myserlis2018} \citep{Komossa2021b}\citep{Komossa2022}. 

Here, we discuss the 2016--2020 outbursts (Fig. \ref{fig:light-CR-Swift}) and deep fades of OJ 287 in context. We then focus on the recent MWL flux and spectral evolution between 2021 and May 2022 that includes the detection of the second-brightest radio flare since the start of the MOMO project. Further, we comment on a recent speculation \citep{Valtonen2021} that a disk impact of a secondary SMBH in OJ 287 might produce detectable precursor flare activity in the optical--UV regime in late 2021 or beyond. 

\section{Outstanding flux and spectral states}

\subsection{2016/2017 and 2020 outbursts} 

With Swift we detected a bright X-ray--UV--optical outburst in 2016--2017 \citep{Komossa2017} \citep{ Komossa2020}. It reached its peak in X-rays in February 2017. These observations motivated a VERITAS (Very Energetic Radiation Imaging Telescope Array System) observation of OJ 287 that was carried out near the X-ray maximum, leading to the first weak detection of VHE (very high energy) emission ($>$ 100 GeV) of OJ 287 at the level of $\sim$ five standard deviations above the background \citep{O'Brien2017}. Surprisingly, in the Fermi $\gamma$-ray band, no accompanying sharp flaring activity was observed.
MOMO radio observations during this epoch reveal that the X-ray--optical outburst is accompanied by a strong radio outburst detected at all frequencies, reaching the highest radio fluxes at high frequencies (up to $\sim$ 11 Jy at 36 GHz) during our ongoing monitoring between 2015 and 2022. 
All previous and new observations clearly establish non-thermal synchrotron emission as the origin of this radio--X-ray flare. 

\begin{figure}
\centering
\includegraphics[clip, width=\columnwidth]{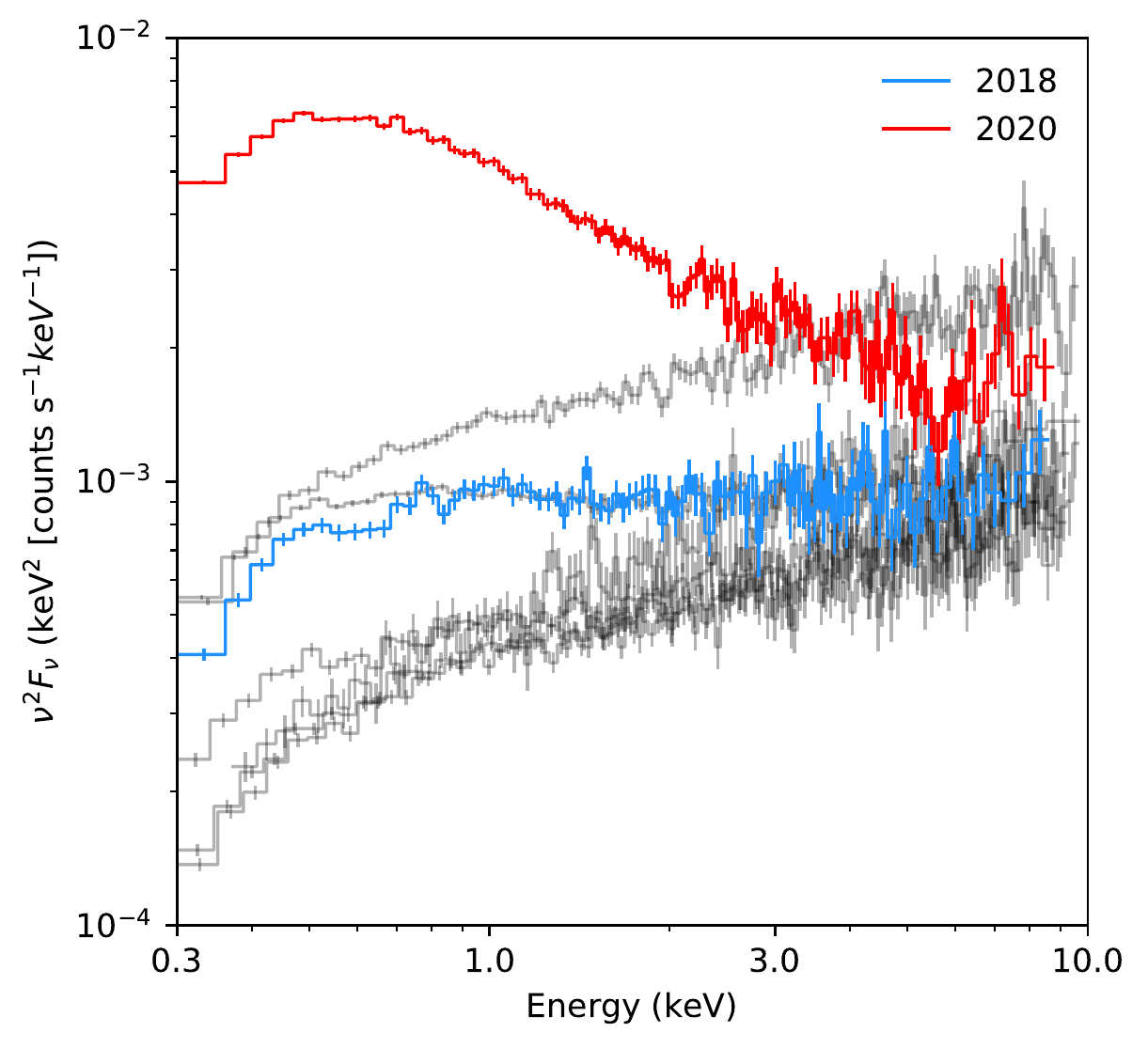}
    \caption{All eight XMM-Newton spectra of OJ 287 taken between 2005 and 2020. Most of the time, OJ 287 is in a relatively flat X-ray spectral state. However, during its 2016/2017 outburst (not covered by XMM-Newton but in our Swift observations) and during its 2020 outburst (red color), an additional supersoft spectral component was discovered that is exceptionally steep for blazars. The XMM-Newton observation obtained quasi-simultaneously with EHT is marked in blue. Adopted from \citet{Komossa2020}. OJ 287 was found in an intermediate flux and spectral state at that epoch. 
    }
    \label{fig:XMM-spec}
\end{figure}

A second bright outburst was detected during the MOMO observations of OJ 287 in 2020 with Swift. XMM-Newton and NuSTAR follow-ups were triggered and caught OJ 287 while it was still near maximum \citep{Komossa2020}. XMM-Newton observations revealed the extremely soft X-ray spectrum of the outburst (Fig. \ref{fig:XMM-spec}). Its synchrotron nature was established through multiple independent arguments including: variability faster than the last stable orbit, assuming an SMBH mass of order 10$^{10}$ M$_\odot$, SED arguments, small optical--UV lags $\le 1$d, and the association with radio flaring.
Along with BL Lac \citep[e.g.,][]{D'Ammando2022}, OJ 287 exhibits the steepest known X-ray spectra among blazars of its kind. 

The 2016/2017 and 2020 outbursts share some similarities with previous, bright, double-peaked, optical outbursts of OJ 287 in the 1970s-1990s \citep[e.g.,][]{Sillanpaa1988, Valtaoja2000}, albeit with lower amplitudes. However, they do not match a pattern of approximately constant separation between the double-peaks of approximately 1 year \citep[0.5-2 yrs; e.g.][]{Valtaoja2000}. 

The 2020 outburst timing is consistent within a few months with predictions of an after-flare predicted by a binary SMBH model of OJ 287 \citep{Sundelius1997}.
In this model, after-flares are produced when new jet activity is launched following a secondary SMBH impact of the disk around the primary, after the disk disturbance has travelled to the inner disk and the accretion rate changes.
However, any such model comes with many assumptions and free parameters, including the physics of the launching of a new jet component, its interactions with the interstellar medium, and the timing when it becomes detectable at multiple frequencies. Alternatively, the 2020 outburst could represent a bright synchrotron flare that is independent of a binary's presence.  

\subsection{2017 deep fade} 

Most of the time, OJ 287 is highly variable with new (lower-amplitude) flares every 1--4 weeks. However, in 2017 we detected with Swift a long-lasting low-state in form of a symmetric fading and recovery of the UV and optical flux that lasted two months \citep{Komossa2021c}, not followed by the X-rays. A previous similar optical deep fade in 1989 was speculated to be perhaps linked to the passing of a secondary SMBH near the jet of the primary SMBH leading to a temporary jet deflection \citep{Takalo1990}. However, using binary model predictions from \citet{Valtonen2021}, the system's geometry in 2017 did not match with a secondary SMBH's location in front of the disk of the primary SMBH at the time of the deep fade. An occultation event by a dusty gas cloud can be ruled out as well, given the lack of UV reddening (and the lack of X-ray absorption). A possible explanation is a temporary jet dispersion or deflection in the core region, or in a  radio-bright off-center quasi-stationary jet feature \citep[e.g.,][]{Hodgson2017, Zhao2022} that was interpreted as either a recollimation shock or a region of maximized Doppler factor in a bent jet. 

\begin{figure}[b]
\centering
\includegraphics[clip, trim=1.0cm 5.3cm 0.9cm 9.0cm, width=\columnwidth]{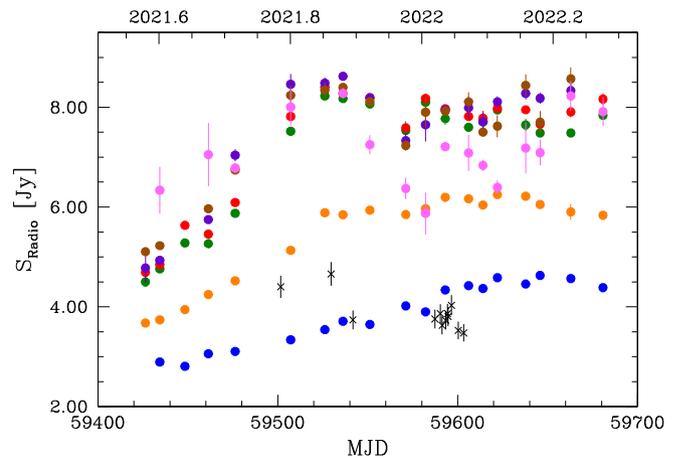}
    \caption{Recent Effelsberg radio observations of OJ 287 obtained in the course of the MOMO project between 2021 August and 2022 May.  During this time interval, the second-brightest radio flare was recorded since the start of the monitoring program, lasting at least until May (colours: blue: 2.6 GHz, orange: 4.85 GHz, green: 10.45 GHz, red: 14.25 GHz, purple: 19.25 GHz, brown: 24.75 GHz, pink: 36.25 GHz, black crosses: SMA data around 230 GHz). 
    }
    \label{fig:radio}
\end{figure}

\subsection{Recent MWL flux evolution and the bright ongoing 2021 -- 2022 radio flare} 

\begin{figure*}
\centering
\includegraphics[clip, trim=0.9cm 5.3cm 1.0cm 2.3cm, width=11.5cm]{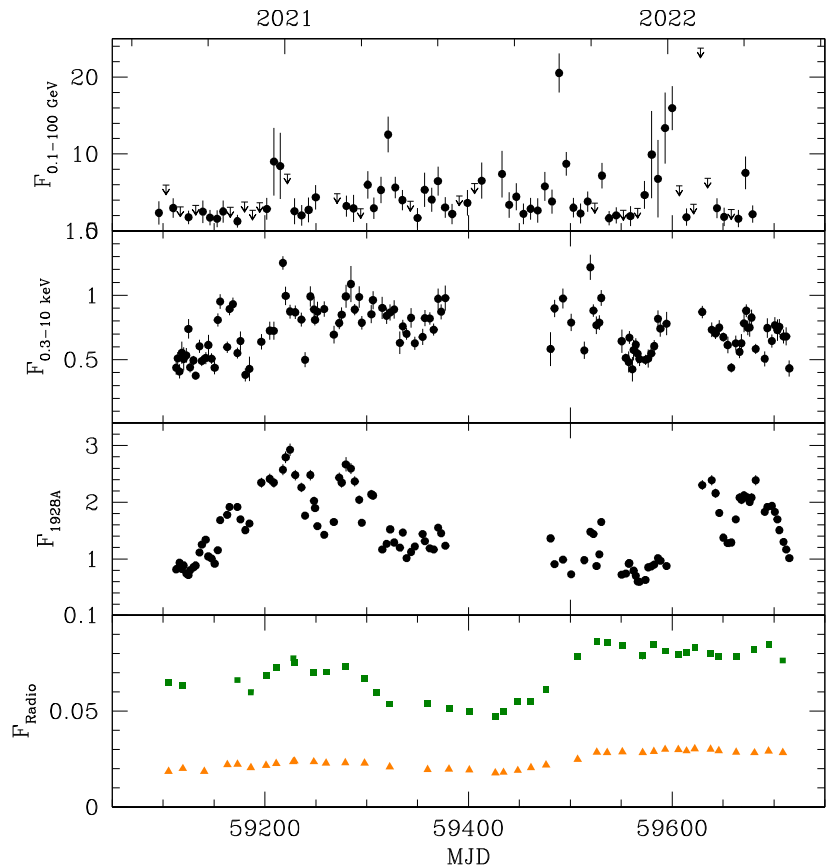}
    \caption{MWL flux light curve of OJ 287 between 2020 September and 2022 May. From top to bottom: Fermi $\gamma$-rays, Swift X-rays and UV (W2 filter at 1928\AA), and Effelsberg radio observations at selected frequencies (orange triangles: 4.85 GHz, green squares: 10.45 GHz).
    The $\gamma$-ray flux (observed 0.1--100 GeV band; one-week averages), the absorption-corrected X-ray flux (observed 0.3--10 keV band), the extinction-corrected UV flux at $\lambda_{\rm obs}$=1928\AA, and the radio flux are in units of 10$^{-11}$ erg s$^{-1}$ cm$^{-2}$. The gap in Swift observations between mid-January to mid-February 2022 is due to a satellite safe mode during which no data were taken. The gap in Swift observations between June to September 2021 is due to the unobservability of OJ 287 with Swift (and from the ground) because of the blazar's close proximity to the Sun. Fermi and Effelsberg observations are not affected by this Sun constraint.   
    }
    \label{fig:MWL}
\end{figure*}

We recorded a strong rise in radio emission in 2021 that peaked in November and then rose again in January (Fig. \ref{fig:radio} at 36 GHz), keeping high flux levels until at least May 2022. This is the second-brightest radio flare since the start of our monitoring in late 2015. Unlike the bright 2016/2017 outburst, the radio is not immediately accompanied by large, long-lasting flaring activity in the optical--UV band where the overall emission level is very low at least until early January 2022 (Fig. \ref{fig:MWL}). 

For the year 2021, the fractional rms variability amplitude $F_{\rm var}$ was computed according to \citet{Vaughan2003} for all radio frequencies between 2.6 and 43.75 GHz, following \citet{Komossa2021c}. It ranges between $F_{\rm var}$=0.105$\pm$0.001 at 2.6 GHz and 0.295$\pm$0.019 at 43.75 GHz. The increase of $F_{\rm var}$ with frequency can be traced back to decreasing opacity. 

Two short, sharp $\gamma$-ray flares of OJ 287 were detected with Fermi in 2021 October and December; the brightest observed in the last few years (Fig. \ref{fig:MWL}). These may be associated with the bright radio flare, matching a pattern reported by \citep{Agudo2011}. On the other hand, the even brighter radio flare of 2016/2017 that was accompanied by the brightest X-ray outburst of OJ 287 so far recorded \citep{Komossa2020} lacked a bright Fermi $\gamma$-ray counterpart. These observations strongly suggest multiple $\gamma$-ray emission sites in OJ 287. 

\section{Recent binary (precursor-flare) activity ?} 

In the context of the binary SMBH scenario, it is conceivable that the interaction of the secondary SMBH with matter of the disk and corona around the primary SMBH contributes emission to the light curve of OJ 287 at select epochs; for instance X-ray emission during epochs of gas accretion. It is interesting to search for such events, though it may be challenging to disentangle them from other flaring that goes on in blazars (and OJ 287 in particular) all the time, completely unrelated to any binary's presence. 
In particular, \citet{Pihajoki2013} and \citet{Valtonen2021} speculated that certain optical lower-amplitude flare activity in the past light curve of OJ 287 that 
was seen to precede ``main flares'' by months was not due to the common blazar variability but was in an unknown fashion related to the binary's presence; for instance through temporary accretion or jet activity of the secondary SMBH, or due to leaky Bremsstrahlung emission from impact-driven streams or bubbles following the secondary's disk crossing. 
They then further suggested that these flares should repeat at similar epochs before new ``main flares'' (assuming that the complicated gaseous physics and magnetic fields involved in all these processes repeat in identical fashion during future secondary disk encounters decades later).    

We have searched our recent MWL light curve of OJ 287 for any such ``precursor flares'', and do not find any evidence for these. 
First, \citet{Komossa2021c} searched for precursor flare activity of the kind suggested by \citet{Pihajoki2013} that was predicted to peak at 2020.96 $\pm${0.1}. No sharp flare (increase by $\sim$2 mag) was observed.   

\begin{figure}[t]
\centering
\includegraphics[width=7.8cm, height=5.3cm]{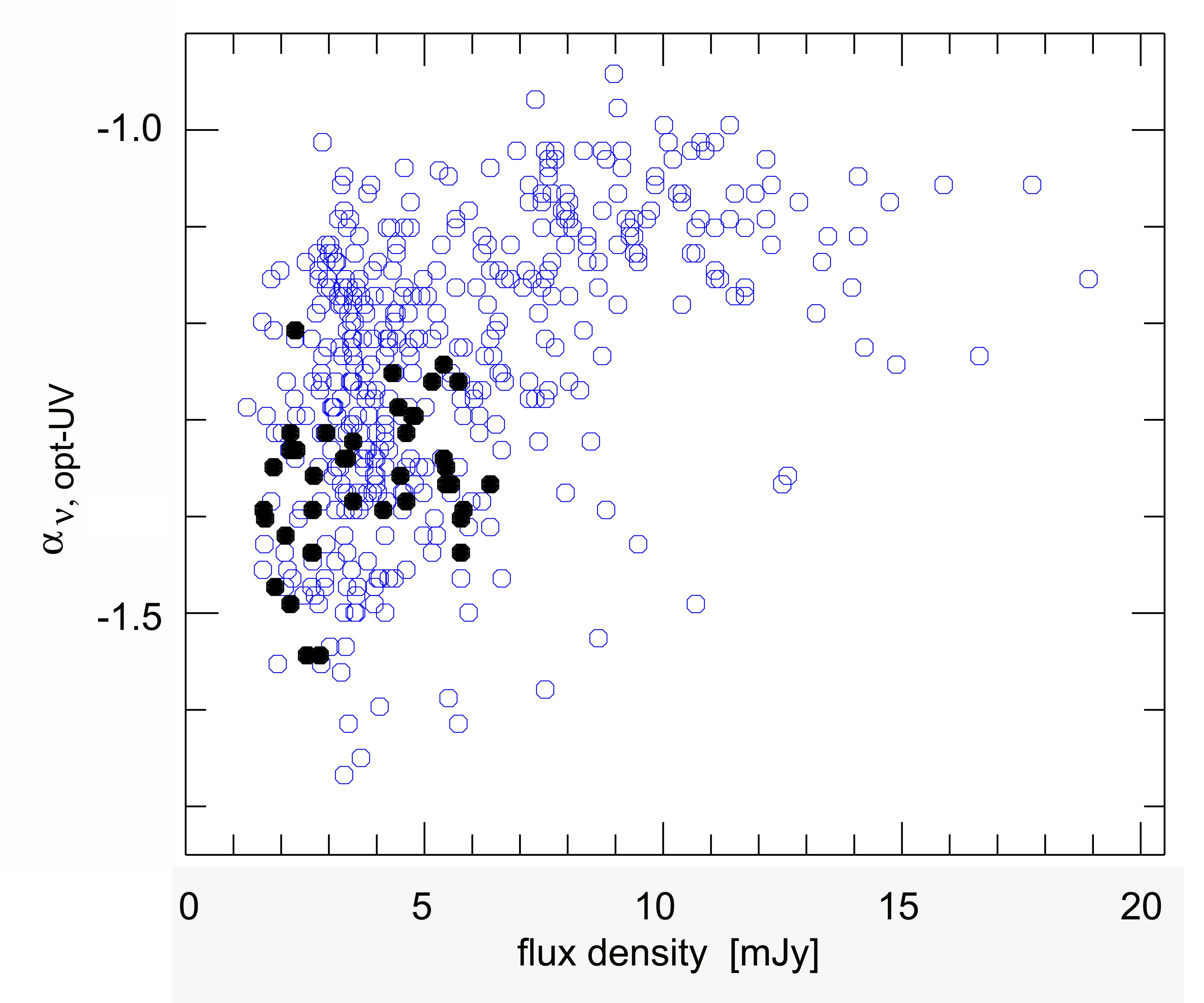}
    \caption{Optical--UV power-law spectral index $\alpha_\nu$ between 2015 December and 2021 November (open circles), and during the period 2021 December to 2022 May (filled circles). The optical flux density $f_{\rm V}$ is measured at $\lambda_{\rm{obs}}$=5468\AA~in units of 10$^{-26}$erg/cm$^2$/s/Hz. OJ 287 displays a "softer-when-brightest" variability pattern. 
    A thermal precursor flare with $\alpha_\nu \sim -0.2$ was predicted to happen in 2021 December. None was detected, neither a flare nor an event with $\alpha_\nu \sim -0.2$. Neither was an event with $\alpha_\nu \sim -0.2$ observed at any other time period. 
    }
    \label{fig:alpha}
\end{figure}

Second, \citet{Valtonen2021} speculated that an optical flare seen in the 2005 light curve of OJ 287 was a precursor flare and would repeat in identical fashion, starting on 2021 December 23 with a thermal Bremsstrahlung spectrum in the optical--UV (and without any X-ray or radio counterpart; their Sect. 3). However, we found OJ 287 in a deep optical--UV low-state throughout December 2021 
and any optical--UV flaring activity at that time period can be excluded (Fig. \ref{fig:MWL}). Further, neither the rise in emission after the deep low-state, nor any other UV--optical flaring activity during the first months of 2022, shows the predicted spectrum of thermal Bremsstrahlung (Fig. \ref{fig:alpha}): 
%
Observed spectral indices $\alpha_{\nu, \rm{opt-UV}}$ (defined as 
$f_\nu \propto \nu^{\alpha_\nu}$, where the optical is measured at 5468\AA~and the UV at 1928\AA) range between
--1.2 and --1.5 while the predicted thermal spectrum would have had $\alpha_\nu \sim -0.2$. 
  
We also note that the pattern of optical-UV variability in late 2021 to early 2022 is very similar to the one in late 2020 to early 2021 (Fig. \ref{fig:MWL}), therefore excluding the possibility that any 2021-2022 flaring activity is driven by non-standard processes that repeated in an identical fashion from 2004-2005.

Third, we find, more generally, that none of the low-amplitude flaring activity of OJ 287 in recent years stands out. All of them, in X-rays, the optical and UV, show the spectral and variability properties that we see in a similar way in our long-term light curves of OJ 287 since 2015 \citep[Fig. \ref{fig:alpha}; see][for our Swift long-term lightcurve]{Komossa2021c}.
Maximum amplitudes of variability in the optical--UV have been rather constant since 2015, with an amplitude of $\sim$1.5 mag in the optical--UV, and variability between 0.1--0.5 cts/s in X-rays, with the single exception of the bright flares in 2016/2017 and 2020. 
We conclude that there is no positive evidence for any binary-driven precursor flare activity in our Swift light curves.  



\section*{Acknowledgments}
SK would like to thank the Swift and XMM-Newton teams for carrying out our observations of OJ 287 and for very useful discussions on the observational set-ups. SK also thanks Ski Antonucci, Mauri Valtonen, and Staszek Zola for enlightening discussions on OJ 287. 
This research has made use of the
 XRT Data Analysis Software (XRTDAS) developed under the responsibility
of the ASI Science Data Center (SSDC), Italy.
This work is partly based on data obtained with the 100-m telescope of the Max-Planck-Institut
f\"ur Radioastronomie at Effelsberg.
The Submillimeter Array near the summit of Maunakea is a joint project between the Smithsonian Astrophysical Observatory and the Academia Sinica Institute of Astronomy and Astrophysics and is funded by the Smithsonian Institution and the Academia Sinica. 
This work made use of data supplied by the UK Swift Science Data Centre at the University of Leicester \citep{Evans2007}.
This work has made use of public Fermi-LAT data 
\citep{Kocevski2021}.
This research has made use of the NASA/IPAC Extragalactic Database (NED) which is operated by the Jet Propulsion Laboratory, California Institute of Technology, under contract with the National Aeronautics and Space Administration.







\end{document}